\documentclass[aps,reprint,amsmath,amssymb,bibnotes,floatfix,microtype,longbibliography]{revtex4-2}

\usepackage{graphicx}
\usepackage[T1]{fontenc}
\usepackage{dcolumn}
\usepackage{bm}
\usepackage[unicode=true,pdfusetitle,
 bookmarks=true,bookmarksnumbered=false,bookmarksopen=false,
 breaklinks=false,pdfborder={0 0 1},backref=false,colorlinks=true]
 {hyperref}
\usepackage{physics}
\usepackage{placeins}
\usepackage{array}
\usepackage{float}
\newcolumntype{M}[1]{>{\centering\let\newline\arraybackslash}m{#1}}

\begin{document}
\preprint{APS/123-QED}

\title{Self-regulated ligand-metal charge transfer upon lithium ion de-intercalation process from LiCoO$_{2}$ to CoO$_{2}$}

\author{Roberto Fantin}
\author{Ambroise van Roekeghem}
\author{Anass Benayad}
\affiliation{Université Grenoble Alpes, CEA-LITEN, 17 rue des Martyrs, 38054 Grenoble, France}%

\date{\today}

\begin{abstract}
Understanding the role of metal and oxygen in the redox process of layered 3d transition metal oxides is crucial to build high density and stable next generation Li-ion batteries. We combine hard X-ray photoelectron spectroscopy and ab-initio-based cluster model simulations to study the electronic structure of prototypical end-members LiCoO$_{2}$ and CoO$_{2}$. The role of cobalt and oxygen in the redox process is analyzed by optimizing the values of d-d electron repulsion and ligand-metal p-d charge transfer to the Co 2p spectra. We clarify the nature of oxidized cobalt ions by highlighting the transition from positive to negative ligand-to-metal charge transfer upon Li$^+$ de-intercalation.
\end{abstract}

\maketitle

\section{INTRODUCTION}

The widespread success of layered lithium transition metal oxides as positive electrode materials in Li-ion batteries is based on their ability to reversibly intercalate Li$^+$ ions and exchange electrons while preserving crystal integrity \cite{liu_understanding_2016}. The archetype for these materials is LiCoO$_2$, introduced three decades ago and still one of the most used cathode materials \cite{manthiram_layered_2021, gent_predicted_2022}. Although LiCoO$_2$ is now regarded as a conventional material in the battery community, the fundamental electron transfer mechanism associated with Li$^+$ de-intercalation and Li-ion cell operation is not yet fully understood, preventing the increase of usable capacity of this material \cite{lyu_overview_2021}. Considering the crystal field splitting of the Co 3d states due to the distorted octahedral coordination, Li$^+$ de-intercalation from LiCoO$_2$ should be compensated by cobalt oxidation from t$_{2g}^6$e$_g^0$ (Co$^{3+}$) to t$_{2g}$$^5$e$_g^0$ (Co$^{4+}$), both in the low-spin configuration. In practice, only about half of lithium ions are typically de-intercalated to avoid fast degradation \cite{lyu_overview_2021}. Such a limit was explained by the Co 3d band pinning to the top of the O 2p one, using qualitative band diagram models \cite{goodenough_li-ion_2013, chebiam_soft_2001}.

In fact, electron withdrawing from the O 2p band has been proposed to trigger deoxygenation and consequent degradation of the layered structure, responsible for performance decrease and eventual failure of the material \cite{dahn_thermal_1994, ensling_nonrigid_2014}. Such behavior is typical for all layered lithium transition metal oxides, of which Li$_x$CoO$_2$ is the forefather member \cite{dahn_thermal_1994, oswald_structural_2023, papp_comparison_2021}. For this reason, Li$_x$CoO$_2$ is taken as a model system in this study, which aims to clarify the fundamental redox mechanism involving both metal and oxygen valence states underlying the performance and limits of this class of materials.

The role of oxygen in the redox process was highlighted in the nineties by density functional theory (DFT) calculations showing that a significant part of the electron transfer happens at the O sites \cite{carlier_first-principles_2003, van_der_ven_first-principles_1998, Zunger_PRL}, explaining the O-O interlayer shrinking observed by \textit{in-situ} X-ray diffraction (XRD) \cite{tarascon_situ_1999}. Including electronic correlations via dynamical mean field theory (DMFT) does not change qualitatively this picture: the average total occupation of the Co 3d shell does not change significantly along LiCoO$_2$, Li$_{0.5}$CoO$_2$, and CoO$_2$ \cite{isaacs_compositional_2020}. However, the linking between DMFT predictions and experimental probes of the electronic structure, notably via spectroscopy techniques, is still missing. Moreover, while oxygen participation to the redox process of LiCoO$_2$ is nowadays accepted in the literature \cite{assat_fundamental_2018, hu_oxygen-redox_2021}, the nature of the co-participating cobalt is doubtful, in particular with respect to the commonly-referred Co$^{3+}$ to Co$^{4+}$ reaction. 

From the structural point of view, the distinction between Co$^{3+}$ and Co$^{4+}$ in Li$_x$CoO$_2$ cannot be easily established. Upon delithiation, the crystal structure of Li$_x$CoO$_2$ is overall preserved even through various phase transitions. These include gliding of the CoO$_2$ layers, distortion of the CoO$_6$ octahedra, and specific Li$^+$ ordering, but only a gradual contraction of the Co-O bond as observed by \textit{in situ} XRD \cite{tarascon_situ_1999, chen_methods_2004, mukai_revisiting_2020}. Even in the case of Li$_{0.5}$CoO$_2$ it is unclear if the low-temperature charge ordered phase displays a complete Co$^{3+}$/Co$^{4+}$ separation \cite{takahashi_single-crystal_2007, motohashi_electronic_2009}. 

X-ray photoelectron spectroscopy (XPS) is a suitable technique to unveil the redox state of cobalt and oxygen in Li$_x$CoO$_2$ since it directly probes the local electronic structure of ions. Previous XPS studies on the deintercalation process of Li$_x$CoO$_2$ suggested that both cobalt and oxygen participate to the redox process, as deduced by the analysis of O 1s and Co 2p XPS spectra based on cluster theory assumptions but without supporting simulation \cite{daheron_electron_2008, ensling_nonrigid_2014}. The Co 2p core level spectra present satellite structures on the high binding energy side, a signature of correlation effects that can be exploited to get deeper insight on the electronic structure of transition metal oxides \cite{zaanen_determination_1986, bocquet_electronic_1996, van_veenendaal_competition_2006}. The local electronic structure of cobalt and oxygen sites was also investigated by X-ray absorption spectroscopy (XAS) \cite{montoro_changes_2000, yoon_oxygen_2002, mizokawa_role_2013}. Mizokawa et al. interpreted the O K-edge XAS spectra changes supported by unrestricted Hartee-Fock density of states calculations for different Co$^{3+}$/Co$^{4+}$ mixtures in the CoO$_2$ triangular lattice. They observed a larger O 2p hole concentration around the Co$^{4+}$ ions \cite{mizokawa_role_2013}. While this interpretation is appealing, it does not explain the satellite structure of XPS spectra. In fact, since the XPS final state is ionized, this technique is more sensitive to the charge transfer satellite structures, revealing the complex interplay within the metal-ligand framework \cite{van_veenendaal_competition_2006, de_groot_2p_2021}. 

To our knowledge, a direct comparison between experimental XPS and theoretical simulations of de-intercalated Li$_x$CoO$_2$, including ligand-metal charge transfer and d-d correlations, is not yet present in the literature. Such aspects are nonetheless critical to characterize the electronic structure of 3d transition metal compounds and quantify the d-d electron repulsion (U$_{dd}$) and the ligand-metal p-d charge transfer energy ($\Delta$) \cite{zaanen_band_1985}.

In this study, we bring new insight on the charge transfer mechanism of Li$_x$CoO$_2$ by combining valence band XPS and core-level Co 2p hard X-ray photoelectron spectroscopy (HAXPES) measurements on thin films electrodes with DFT-based single cluster model calculations for the end-members LiCoO$_2$ and CoO$_2$. We find that delithiation drives the compound from the mixed valence regime in LiCoO$_2$ to the negative charge transfer regime in CoO$_2$, leaving the net number of electrons in the Co 3d shell nearly constant. Yet, we observe a reorganization of the electronic structure: the de-lithiation process is compensated by an electron withdrawal from the t$_{2g}$ states while there is an electron density backflow from O 2p to the e$_g$ states.

\section{RESULTS AND DISCUSSION}

\subsection{Transition from positive to negative charge transfer upon de-lithiation}

To set the basis for our cluster model Hamiltonian, we first obtained a p-d tight binding model from the wannierization of converged paramagnetic DFT calculations of LiCoO$_2$ and CoO$_2$ electronic structures. Therefore, we reduced at minimum the dependence of our model to semi-empirical parameters, in particular the on-site energies and hopping parameters, to emphasize the role of U$_{dd}$ and $\Delta$ on the electronic structure of LiCoO$_2$ and CoO$_2$. The hopping terms for the CoO$_6$ cluster were extracted from the wannier tight binding model. The Hamiltonian was then augmented with the Coulomb interaction and spin-orbit coupling of the 3d shell, while the charge transfer was treated by means of configuration interaction model.

\begin{figure}[hbt!]
\includegraphics[width=.95\linewidth]{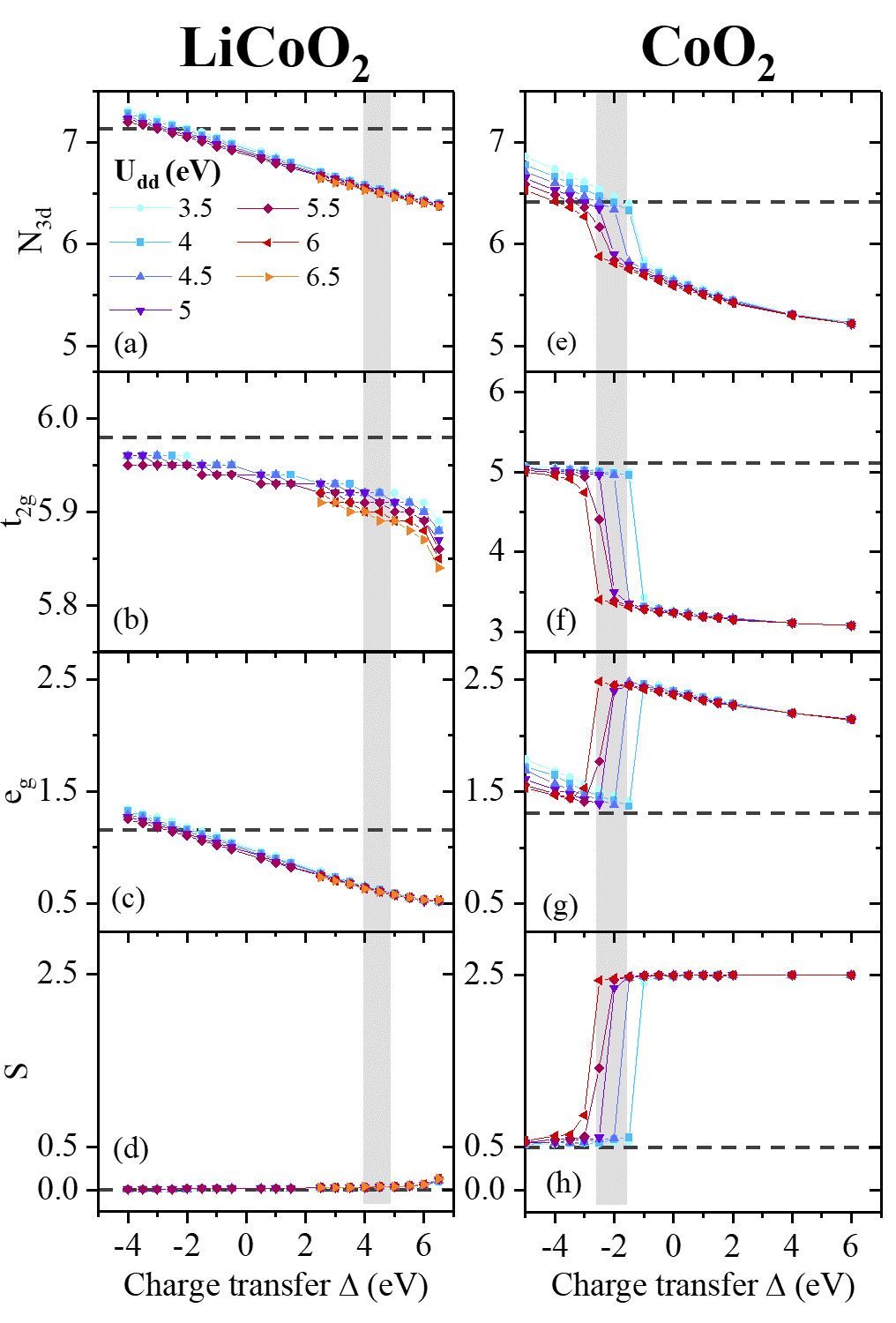}
\caption{\label{fig1} Ground state character by cluster model calculations of (left panels) LiCoO$_2$ and (right panels) CoO$_2$: (a,e) total Co 3d, (b,f) t$_{2g}$, and (c,g) e$_{g}$ electronic occupations and (d,h) total spin S as a function of U$_{dd}$ and $\Delta$. The dashed lines indicate the values obtained by exact diagonalization of the p-d tight binding model only. The gray areas indicate the range where we observed best agreement with our photoelectron spectroscopy measurements.}
\end{figure}

\begin{table}[hbt!]
\caption{\label{tab:table1} Hubbard U$_{dd}$ and charge transfer $\Delta$ energies for LiCoO$_2$ evaluated by (a) semi-empirical cluster model calculations and (b) linear response approach, and (c) constrained random phase approximation, compared to our results. The values in parenthesis are for CoO$_2$.}
\begin{ruledtabular}
\begin{tabular}{c c c}
U$_{dd}$ (eV) &
$\Delta$ (eV) &
\textrm{Reference}\\
\colrule
3.5                     & 4                     & \cite{van_elp_electronic_1991} $^{(a)}$ \\
6.5                     & 1                     & \cite{mizokawa_role_2013, ikedo_electronic_2010} $^{(a)}$ \\
5.5                     & -0.5                  & \cite{mesilov_charge_2013} $^{(a)}$ \\
4.91 (5.37)             & --                    & \cite{zhou_first-principles_2004} $^{(b)}$\\
4.40 (3.68)             & --                    & \cite{kim_quantification_2021} $^{(c)}$ \\
4.5 $\pm$ 1.0 & 5.0 $\pm$ 1.5 & This work \\
(4.0 $\pm$ 0.5) & (-2 $\pm$ 0.5) & \\
\end{tabular}
\end{ruledtabular}
\end{table}

The ground state for the cluster model Hamiltonian was obtained by exact diagonalization method as implemented in \textsc{Quanty} \cite{haverkort_multiplet_2012}, leaving only U$_{dd}$ and $\Delta$ as empirical parameters. In all our calculations, we restricted the Hamiltonian to evaluate configurations between d$^n$ and d$^8$ with n = 6 for LiCoO$_2$ and n = 5 for CoO$_2$. Figure \ref{fig1} shows the expected values of the number of electrons in the Co 3d, t$_{2g}$, and e$_g$ shells and the total spin S as a function of U$_{dd}$ and $\Delta$. In each panel, the dashed line shows the value obtained by the solution of the tight binding Hamiltonian only. 

For LiCoO$_2$, the occupation of the Co 3d shell estimated by this method is larger than the nominal value of six, as expected for the Co$^{3+}$ oxidation state (Fig. \ref{fig1}a). Specifically, while the t$_{2g}$ states are almost filled (Fig. \ref{fig1}b), an occupation of about 1 is observed for the e$_g$ orbitals (Fig. \ref{fig1}c). This is a clear effect of the strong covalence between Co 3d and O 2p orbitals, recalling that the e$_g$ orbitals point towards the ligand O 2p ones in the (distorted) octahedral crystal field. By introducing electronic correlations in our many-electrons cluster model, we observe a net decrease of the Co 3d total occupation. Nevertheless, the cluster calculation is more influenced by $\Delta$ than by U$_{dd}$, in particular regarding the e$_g$ occupations (Fig. \ref{fig1}c). With low $\Delta$, the displacement of electrons from O 2p to Co 3d is energetically favored; in the case of a negative $\Delta$, this process dominates against the intra-atomic Coulombic repulsion.

To find out which parameters represent better the electronic structure of LiCoO$_2$, we compared the simulated and HAXPES Co 2p spectra \cite{Suppl_Inf}. The gray area in the panels indicate the range of values that better satisfied these conditions and we compare our result with other values of U$_{dd}$ and $\Delta$ in Table \ref{tab:table1}. In the literature, no agreement is found regarding the electronic structure of LiCoO$_2$, which was defined as intermediate \cite{van_elp_electronic_1991}, charge transfer \cite{ikedo_electronic_2010} or even negative charge transfer insulator \cite{mesilov_charge_2013}, while from DFT calculations LiCoO$_2$ is a band insulator, due to the complete filling of the t$_{2g}$ band \cite{isaacs_compositional_2020}.

Our simulations highlight that the dominant aspect in the electronic structure of LiCoO$_2$ is the Co 3d - O 2p covalence, already described by DFT and subsequently corrected to better describe the effects of correlations. This allowed us to classify LiCoO$_2$ as a mixed-valence phase following the modern classification for high-valence transition metal oxides \cite{sawatzky_g_2016}, in line with van Elp et al. \cite{van_elp_electronic_1991}. To get a general guideline for U$_{dd}$, the screened repulsion values evaluated by first-principles approaches are also reported in the Table and in line with our findings. 

Regarding the electronic structure of CoO$_2$, the values U$_{dd}$ $\approx$ $\Delta$ $\approx$ 4.5 eV obtained for LiCoO$_2$ gave a reasonable starting point from which we followed the trends identified by Bocquet et al. \cite{bocquet_electronic_1996}: with increasing oxidation state, U$_{dd}$ slightly increases while $\Delta$ abruptly decreases, because of orbital shrinking and larger electronegativity. The DFT-based tight binding solution of CoO$_2$ shows a slightly lower Co 3d occupation than in LiCoO$_2$, although still larger than six electrons (Fig. \ref{fig1}e). While the t$_{2g}$ shell lost about one electron, the occupation of the e$_{g}$ orbitals even increased (Fig. \ref{fig1}f,g). The system was found to be in low spin configuration, in qualitative agreement with experimental data \cite{motohashi_synthesis_2007}.

In the cluster calculations, with decreasing $\Delta$, we observed a transition at $\Delta$ $ \approx$ -2 eV from cobalt high-spin (t$_{2g}^3$e$_g^2$, S = 5/2, HS) to low spin (t$_{2g}^5$e$_g^1$, S = 1/2, LS) configuration. The best agreement with the experimental spectra was obtained for values just below such transition (Tab. \ref{tab:table1}). As a direct consequence of the clear negative charge transfer nature of this compound, the electronic structure in the low spin region is not in the t$_{2g}^5$e$_g^0$ configuration usually referred to Co$^{4+}$ oxidation state in the literature.

This result gives another perspective on the role of oxygen in the charge transfer mechanism. The O 2p participation directly results from the increased hybridization between the e$_g$ and O 2p states while one net electron is extracted from the t$_{2g}$ band. The oxygen atoms coordinated around cobalt are all equally involved in this process, with an average hole concentration increasing from 0.09 in LiCoO$_2$ to 0.23 in CoO$_2$ per oxygen atom in the cluster. Moreover, for CoO$_2$, the electronic occupations obtained by our cluster model are similar to our DFT predictions and to DFT+DMFT results \cite{isaacs_compositional_2020}. 

\subsection{Valence band analysis}

\begin{figure}[hbt!]
\includegraphics[width=.95\linewidth]{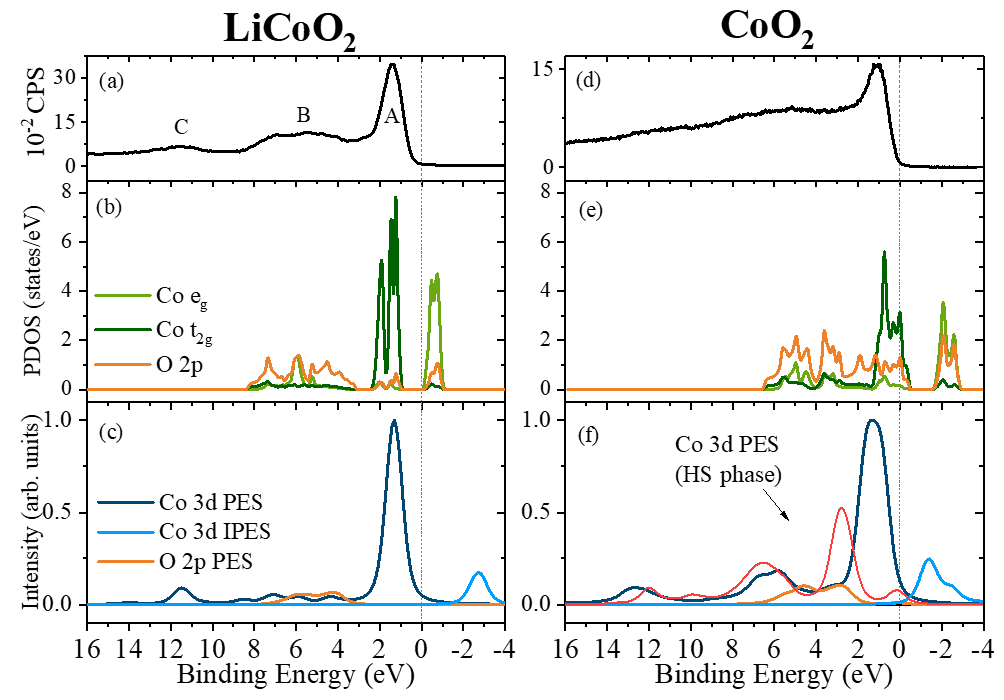}
\caption{\label{fig2} Comparison of LiCoO$_2$ and CoO$_2$ (a,d) XPS valence bands, (b,e) DFT PDOS, and (c,f) electron removal (PES) and addition (IPES) spectral function. The PDOS for LiCoO$_2$ in panel b was shifted to match with the main experimental peak. The O 2p spectral functions were obtained by averaging the curves obtained from all O atoms in the cluster and scaling them to the actual composition and to the photoelectron cross section $\sigma$ ($\sigma_{Co_{3d}}$ / $\sigma_{O_{2p}}$ $\approx$ 10 \cite{leckey_subshell_1976}). The U$_{dd}$ and $\Delta$ parameters for LiCoO$_2$ (LS CoO$_2$) are 4.5 and 4.5 eV (4.5 and -2 eV). For the HS CoO$_2$ phase, $\Delta$ = 4.5 eV.}
\end{figure}

Figure \ref{fig2} shows the experimental XPS valence band of the (a) pristine LiCoO$_2$ and (b) cycled Li$_{0.12}$CoO$_2$ thin films. The valence band of LiCoO$_2$ presents a narrow peak at 1-2 eV (A) followed by a large band between 3 to 8 eV (B) and a smaller one at 11-12 eV (C), in accordance with the literature \cite{daheron_electron_2008, ensling_nonrigid_2014, galakhov_electronic_2002}. For the cycled Li$_x$CoO$_2$, we note that the contribution of F 2p, and O 2p from surface species deposed after Li$^+$ cycling cannot be neglected \cite{dedryvere_xps_2005}, although the main valence band contribution is to be assigned to Li$_x$CoO$_2$. 

The experimental valence spectra are compared to the DFT partial density of states (PDOS) and the Co 3d and O 2p electron removal spectra for LiCoO$_2$ and CoO$_2$. For LiCoO$_2$, the DFT PDOS fit well with features A and B, but do not reproduce the small feature C at higher binding energies (Fig. \ref{fig2}b). In contrast, the cluster calculation (Fig. \ref{fig2}c) allows to reproduce this peak, assigned to a charge transfer satellite, while preserving the overall structure of the PDOS. Features A and C are therefore related to the screened and unscreened t$_{2g}$ photoelectrons, respectively. Their relative intensity is proportional to $\alpha$ and $\beta$ in the 1-electron removal wave function, expressed in the configuration interaction framework of our model as $\Psi = \alpha\ket{d^n} + \beta\ket{d^{n+1}\boldsymbol{L}} + \gamma \ket{d^{n+2}\boldsymbol{L^2}}$, where $\boldsymbol{L}$ denotes a hole in the ligand shells ($\gamma$ being negligible). Feature B is related to O 2p and e$_g$ mixed orbitals, as evident from both the PDOS and the electron removal spectra. For the CoO$_2$ valence band, we observe an increasing mixing between the O 2p and Co 3d states and a closing of the gap between the two bands (Figs. \ref{fig2}e,f), which follows the experimental observation. Finally, we note that the simulated Co 3d spectra for the HS phase does not match with the experimental valence band, indicating that the cobalt ions are indeed in LS state. 

\begin{figure}[hbt!]
\includegraphics[width=.95\linewidth]{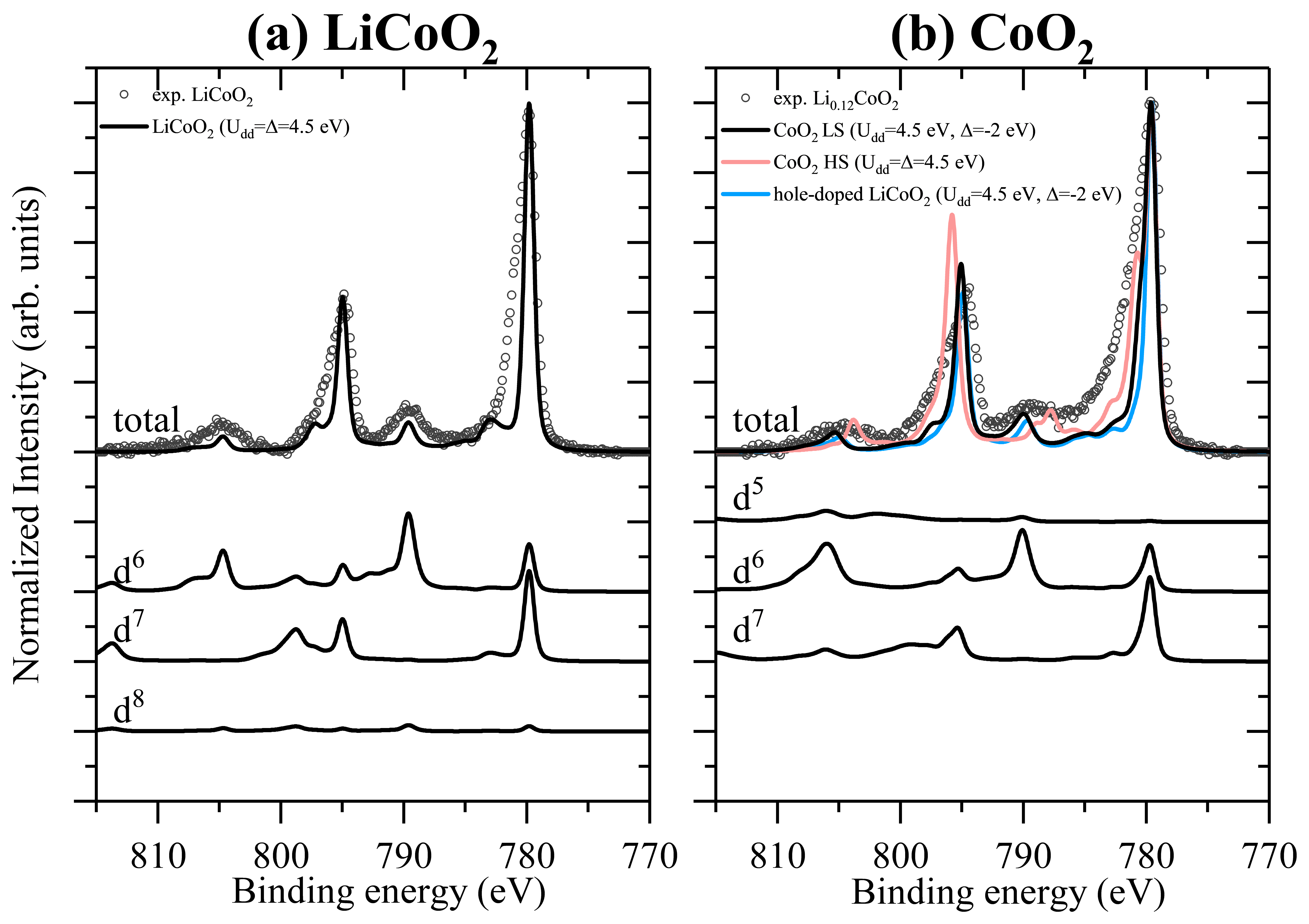}
\caption{\label{fig3} (a,b) Co 2p core level photoemission spectra computed within a cluster model and compared to in-lab HAXPES experimental data with h$\nu$ = 5.4 keV. The depth sensitivity was estimated with the TPP2M method as $\sim$ 15 nm \cite{tanuma_calculations_1994, fantin_revisiting_2022}. The experimental data is shown as empty circles after background subtraction of an iterated Shirley function. All simulated spectra were normalized and shifted to the experimental Co 2p$_{3/2}$ peak maximum.}
\end{figure}

\subsection{Insight from Co 2p HAXPES satellite peaks}

To simulate the core photoemission spectra, the ground state Hamiltonian was augmented with Co 2p core level including core-valence multiplet interactions and spin-orbit coupling. The resulting spectra obtained with representative values for U$_{dd}$ and $\Delta$ are compared to experimental hard X-ray photoelectron spectroscopy (HAXPES) Co 2p spectra in Fig. \ref{fig3}. These were obtained using a Cr K$\alpha$ X-ray source (5.4 keV) to reduce the contribution of the uppermost surface contamination (pristine) or degraded (cycled) surface layers and overcome the overlapping with Co LMM Auger transitions \cite{fantin_revisiting_2022}. As shown in Figs. \ref{fig3}a,b the Co 2p$_{3/2}$ and Co 2p$_{1/2}$ spin orbit components are separated by about 15 eV and are each constituted by a mainline ($\sim$ 780, 795 eV) and a satellite ($\sim$ 790, 805 eV) peak. This structure, typical for 2p core photoemission in transition metal oxides, has been explained in the framework of cluster model theory, assigning the mainline and satellite peaks to the locally screened ($\ket{2p^5d^{n+1}\boldsymbol{L}}$) and unscreened ($\ket{2p^5d^{n}}$) states, respectively \cite{van_veenendaal_competition_2006, hufner_photoelectron_2003}. Our simulated Co 2p spectrum of LiCoO$_2$ (Fig. \ref{fig3}a) matches with this interpretation, so far only assumed in the literature \cite{daheron_electron_2008, ensling_nonrigid_2014, galakhov_electronic_2002}. To get more insight into the core spectra, we also computed partial spectra by applying restrictions to the configurations considered by the transition operator. These highlight that d$^6$ and d$^7$ are the main contributions (56\% and 38\%, respectively), in agreement with the ground state electronic occupations (Figs. \ref{fig1}a-e). The spectral distributions for these configurations agree with the interpretation given in literature: while a mixed contribution is observed in the main line, the satellite peak is uniquely present in the d$^6$ partial spectrum. The simulation fits overall well with the experimental data except for the asymmetry of the main line. This has been referred to non-local charge transfer screening processes, which indeed are not included in our single cluster model but can be obtained by either multicluster or DMFT calculations, as shown for other transition metal oxides \cite{daheron_electron_2008, ghiasi_charge-transfer_2019, van_veenendaal_nonlocal_1993, taguchi_revisiting_2008}. The larger experimental intensity between the asymmetric main line and the satellite ($\sim$ 785 eV) can be referred instead to Co$^{2+}$ ions formed by surface degradation typically associated to battery cycling and eventually visible even with HAXPES measurements \cite{fantin_revisiting_2022}. 

The Co 2p HAXPES spectrum for the deeply delithiated Li$_{0.12}$CoO$_2$ thin film shows two major changes from pristine LiCoO$_2$, namely the broadening of the main line and an increase in intensity around 785 eV. In the literature, both ex-situ and operando HAXPES experiments for delithiated Li$_{x}$CoO$_2$ related the broadening of the main peak to the Co$^{3+}$/Co$^{4+}$ oxidation \cite{daheron_electron_2008, ensling_nonrigid_2014, operando_HAXPES}. The absence of a new satellite for Co$^{4+}$ is a consequence of the low weight of a d$^5$ electronic configuration. The simulated Co 2p spectra (fig. \ref{fig3}b) of CoO$_2$ assigned to t$_{2g}^5$e$_g^1$ electronic structure (fig. \ref{fig1}), present similarities to LiCoO$_2$, without significant change of the satellite position. The ground state of CoO$_2$ has a mixed character with a majority weight for the d$^6$ configuration (10\%, 45\% and 38\% for d$^5$, d$^6$ and d$^7$ configurations, respectively), due to the increase in Co 3d - O 2p hybridization. Because of such distribution, the d$^6$ satellite peak is still present while the contribution from d$^5$ configuration is negligible. Note that the partial spectra in Fig. \ref{fig3}b refers to the total electron number in the 3d shell, however we observe a reorganization between t$_{2g}$ and e$_{g}$ in our ground state calculations (Figs. \ref{fig1}f,g). The asymmetry of the main line can be related to the non-local screening channel due to the metallic character of CoO$_2$. For the sake of comparison, in Fig. 3b we show the simulated spectra for a HS configuration (red line), in which the d$^5$ contribution dominates the spectra as shown in fig. \ref{fig1}e. 

Finally, to understand the correlation between crystal structure change from O3 to O1 (using Delmas' notation \cite{delmas_structures}) and the negative charge transfer transition, we performed a calculation using the LiCoO$_2$ cluster model with one extra hole and the same parameters obtained for CoO$_2$. The resulting Co 2p spectrum (blue line), agrees well with the one obtained by the formal O1 CoO$_2$ structure, suggesting that the electronic structure reorganization is not mainly driven by the small structural changes observed upon de-lithiation.

A further generalization can be made considering the cobalt local electronic structure in the perovskite SrCoO$_3$. A similar trend for the Co 2p spectra was observed in the La$_{1-x}$Sr$_{x}$CoO$_{3}$ ($0<x<1$) system, where the formal oxidation state of octahedrally coordinated Co also goes from Co$^{3+}$ (x=0) to Co$^{4+}$ (x=1) but a negative charge transfer state with intermediate spin (t$_{2g}^4$e$_g^2$, S=3/2) was instead predicted for the SrCoO$_3$ end-member by cluster model simulations \cite{Potze_IS_SrCoO3, Saitoh_IS_LaSrCoO3}. In our calculations, such state was obtained for U$_{dd}$ = 5.5 eV and $\Delta$ = -2.5 eV within the high-to-low spin transition region (fig. \ref{fig1}h). Such configuration was however excluded by our Co 2p HAXPES analysis \cite{Suppl_Inf} supporting the LS configuration for this compound. 
The difference in the local Co 3d structure between SrCoO$_3$ and CoO$_2$ can be related to the different connectivity of CoO$_6$ octahedra and crystal field, leading to the different magnetic behavior observed experimentally \cite{Long_IS_SrCoO3, motohashi_synthesis_2007}. In both cases, however, the Co$^{4+}$ formal oxidation state appears instead to be stabilized by negative charge transfer from surrounding O 2p states. 

\section{CONCLUSIONS}

In summary, we found a reorganization in the local electronic structure of CoO$_2$ driven by the decrease of the charge transfer energy towards negative values. As a negative charge transfer material, the electronic structure of CoO$_2$ is better described as 3d$^6\boldsymbol{L}$, in which the charge extracted from the t$_{2g}$ states is balanced by a backflow from O 2p to e$_g$ orbitals. The decrease of both O 2p and t$_{2g}$ electron occupations finds good agreement with the published O K-edge XAS as well as resonant inelastic X-ray spectroscopy (RIXS) studies \cite{hu_oxygen-redox_2021, mizokawa_role_2013}, which however did not highlight the non-zero occupation of the e$_g$ states.

This combined HAXPES and ab-initio-based cluster model simulations study shows the importance of considering electronic correlations and charge transfer theory to interpret the XPS/HAXPES core level spectra and understand the redox process of layered 3d transition metal oxides. Our results highlight the fundamental role of self-regulated ligand-metal charge transfer explaining the so-called rehybridization mechanism \cite{zhang_pushing_2022, Zunger_PRL} in a quantitative manner and based on experimental measurements, which reveal substantial impact on both cobalt and oxygen local electronic structure. 

It is forecasted that the redox process in all next-generation cathode materials are based on such underlying mechanism. Indeed, our results come in a context where understanding the role of oxygen in transition metal oxides is considered to be a key objective. Similarly to Li$_x$CoO$_2$, oxygen-driven redox mechanisms have been proposed for Li-rich and Ni-rich oxides targeted next-generation positive electrode materials, for which the role of oxygen is nowadays under intense discussion as recently reviewed in ref. \cite{zhang_pushing_2022}. Noteworthy, recent studies on the parent material Li$_x$NiO$_2$ proposed a central role of negative charge transfer in the charge compensation mechanism starting from the lithiated compound \cite{foyevtsova_linio_2019, menon_oxygen-redox_2023, genreith-schriever_oxygen_2023}: this perfectly complements our study for Li$_x$CoO$_2$, in line with the general trend going from early to late 3d transition metals \cite{sawatzky_g_2016}. 

\section{METHODS}

\subsection{Electrochemical Li$^+$ de-intercalation of LiCoO$_2$ thin films}

We refer to previous publications for details on the preparation and characterization of the pristine LiCoO$_2$ thin films \cite{fantin_revisiting_2022, oukassi_millimeter_2019}. The Li$^+$ deintercalation was performed electrochemically in a homemade Li-ion cell designed for the LiCoO$_2$ thin film electrodes. The cell case in polyether ether ketone (PEEK) consists of a bottom part with a cavity for the cell components stack and two gasket rings to ensure air-tightness and a cover with the two current collector tips. The LiCoO$_2$ electrodes of active surface area 295.5 mm$^2$ were cut from the SiO$_2$ wafer using a diamond knife to fit in the cavity of the cell case (34 x 26 mm$^2$). The cell components are listed in order of assembling: LiCoO$_2$ thin film onto Pt substrate, Viledon and Cellgard separators, 300 µL of a 1 M LiPF$_6$ solution in ethylene carbonate (EC) and ethyl methyl carbonate (EMC) with weight ratio 3:7 (LP57 electrolyte, Sigma Aldrich), and Li foil (Rockwood, 135 $\mu$m) with area larger than the active area of LiCoO$_2$, and a SS disk interposed between Li and the current collector tip. To avoid electrochemical corrosion, both negative and positive current collectors were protected by Al foil, respectively. After 10 hours at open circuit voltage, the cell was galvanostatically charged with 73 $\mu$A of current, corresponding to a C-rate of C/100 assuming a total capacity of 270 mAh/g. Due to internal corrosion at high voltage, we stopped the deintercalation at 4.8 V vs Li$^+$/Li. According to the LiCoO$_2$ deintercalation curve, this corresponds to about 90 \% Li$^+$ extraction \cite{amatucci_coo2_1996}.

To confirm the stoichiometry of the pristine and cycled thin films, they were characterized by ICP-MS. Three samples (between 50 and 100 mg) were cut with a diamond knife from the electrodes and dissolved in 3 mL of a 65 \% aqueous solution of HNO$_3$ and 5 mL of a 30 \% aqueous solution of HCl. The dissolution was supported by a microwave treatment (Multiwave 3000, Anton Paar) at 800 W for 35 minutes, repeated twice after 2 mL of ultrapure water between the two steps. The Si substrate was not attacked by the treatment. The Li and Co concentration of the recovered solutions were measured with the ICPMS 7900x (Agilent Technologies) spectrometer with the following conditions: RF power 1500 W, plasma Ar gas flow rate of 15 L/min, nebulizer Ar gas flow rate of 1.15 L/min, and integration time of 0.1 s. The average Li/Co atomic concentration percentage ratio for the pristine and de-intercalated film are 97 $\pm$ 2 \% and 12 $\pm$ 3 \%, respectively. 

The crystal structure of the cycled LiCoO$_2$ thin film was investigated by X-ray Diffraction (XRD) using a Bruker D8 Advance equipment employing a Cu K$\alpha$ X-ray tube. The sample was encapsulated by Kapton tape to avoid structural degradation induced by air contamination. The diffractogram matched with the CoO\textsubscript{2} O3 structure (PDF 04-015-9980), consistently with the uncompleted deintercalation. For this reason, we performed the simulations for CoO\textsubscript{2} with both the O1 and O3 structures. The two starting structures gave a very similar result in terms of ground state electronic structure and XPS simulations. 

\subsection{Soft and hard X-ray photoelectron spectroscopy}

XPS and HAXPES measurements were performed with a QUANTES spectrometer (ULVAC-PHI) equipped with a co-localized dual X-ray source consisting of monochromatic Al K$\alpha$ (1486.6 keV, 25 W) and Cr K$\alpha$ (5414.9 keV, 50 W) sources. The samples were transferred from the Ar-filled glovebox to the XPS chamber using a dedicated airtight transfer vessel. The X-ray beam spot size was 100 $\mu$m diameter and the take-off angle for photoelectron detection was 45°. The experiments were performed under ultrahigh vacuum conditions (p < 10\textsuperscript{-7} Pa). High-resolution spectra were acquired with a pass energy of 69 eV for both energy sources, corresponding to an energy resolution of 0.81 (Al K$\alpha$) and 1.16 eV (Cr K$\alpha$), as estimated from by the FHWM of the Ag 3d\textsubscript{5/2} of a reference Ag sample. Automatic double charge neutralization was employed for the pristine LiCoO\textsubscript{2} film but not for the deintercalated one, exploiting the insulator-to-metal transition of Li\textsubscript{x}CoO\textsubscript{2} \cite{motohashi_electronic_2009}. Binding energy charge correction was performed using the Fermi level of the Pt sublayer for the LiCoO\textsubscript{2} thin film. An iterated Shirley background was subtracted using the CasaXPS software \cite{fairley_systematic_2021}.

\subsection{DFT calculations and wannierization}

The electronic structure of LiCoO\textsubscript{2} and CoO\textsubscript{2} was calculated with the full-potential linearized augmented-plane-wave method of \textsc{Wien2k} \cite{blaha_wien2k_2020}, using the Perdew-Burke-Ernzerhof (PBE) generalized gradient approximation (GGA) for the exchange-correlation potential. The atomic muffin-tin sphere radii for Li, Co, and O were 1.77, 1.94 (1.9), and 1.67 (1.63) for LiCoO\textsubscript{2} (CoO\textsubscript{2}). The plane wave cutoff was set to R\textsubscript{MT}\textsuperscript{min}K\textsubscript{max }= 7, with R\textsubscript{MT}\textsuperscript{min } the smallest atomic sphere radius and K\textsubscript{max} the largest k-vector. The paramagnetic ground state electronic density was obtained with a 10x10x10 wavevector-grid in the primitive full Brillouin zone with convergence criteria of 1.36 meV/f.u. for the total energy and 10\textsuperscript{-3} electrons/f.u. for the charge, respectively. The crystal structure of LiCoO\textsubscript{2} (R-3m, a=2.82 {\AA} and c=14.05 {\AA}) and O1-CoO\textsubscript{2} (P-3m1, a=2.82 {\AA} and c=4.29 {\AA}) were taken from ref \cite{amatucci_coo2_1996}. The structure of O3-CoO2 was taken from ref. \cite{isaacs_compositional_2020}.

\begin{figure}[hbt!]
\includegraphics[width=\linewidth]{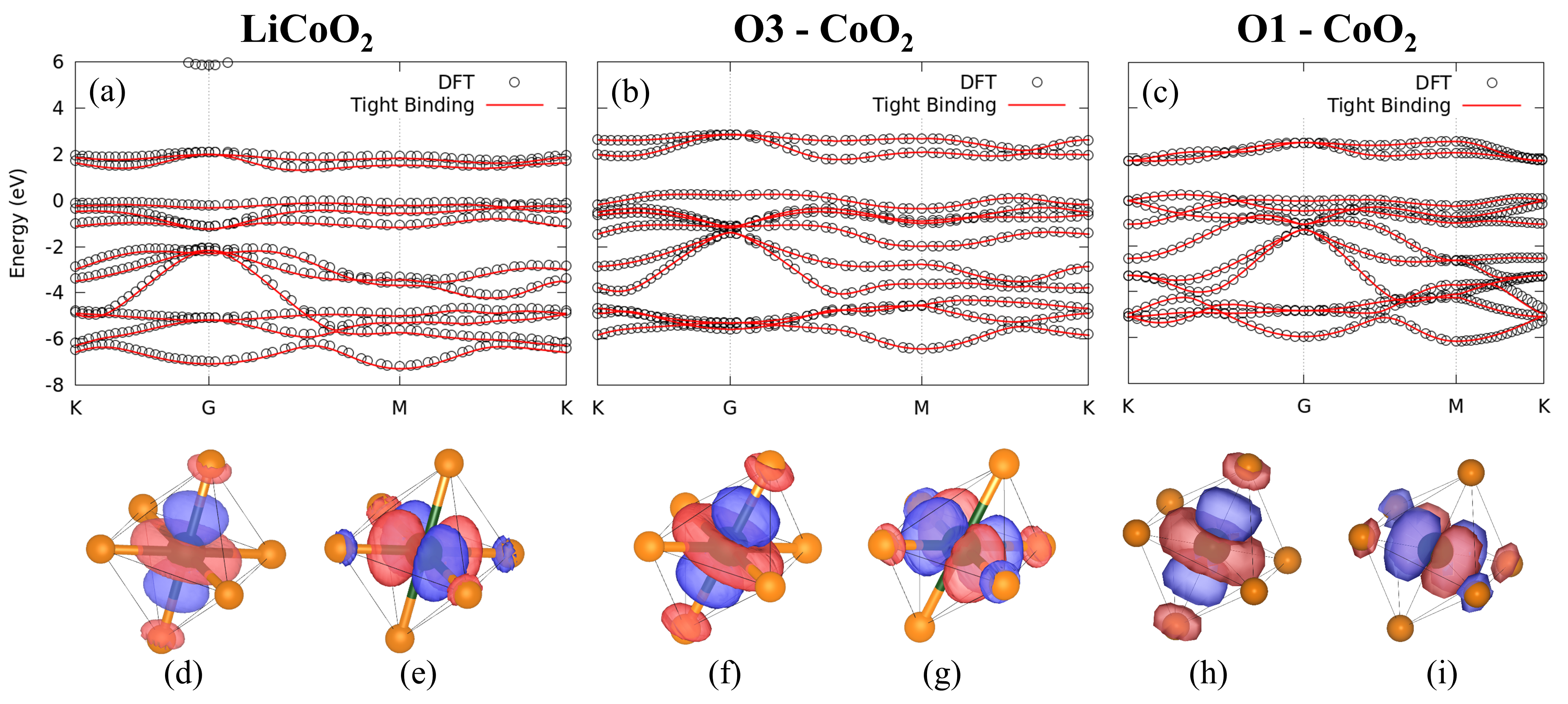}
\caption{\label{fig4} Interpolated band structures of (a) LiCoO$_2$, (b) O3-CoO$_2$ and (c) O1-CoO$_2$. The real-space plots of the (d,f,h) d\textsubscript{z$^2$} and (e,g,i) d\textsubscript{x$^2$-y$^2$} MLWFs are shown below the respective input band structure. Green and orange spheres indicate cobalt and oxygen atoms, respectively.}
\end{figure}

The converged electronic structures were used to create Maximally Localized Wannier Functions (MLWFs) using \textsc{Wannier90} through the \textsc{Wien2wannier} package \cite{pizzi_wannier90_2020, kunes_wien2wannier_2010}. We kept the same number of k-points and restrict to the p-d low energy range (between -8 and 4 eV), leading to 11 MLWFs for both LiCoO\textsubscript{2} and CoO\textsubscript{2}. To minimize the off-diagonal intra-atomic hopping terms, the reference system was rotated to align with the CoO\textsubscript{6} (distorted) octahedra. The wannierization was performed with a convergence criteria for the totals spread of 10\textsuperscript{-10} Å\textsuperscript{2}. The interpolated band structures superimposed to the DFT results are shown in figure \ref{fig4}a-c. The MLWFs were plotted in real space with the \textit{wplot} tool of \textsc{wien2wannier} using a 80x80x80 grid over a 3x3x3 supercell. The d\textsubscript{z$^2$} and d\textsubscript{x$^2$-y$^2$} MLWFs of LiCoO\textsubscript{2}, O3-CoO\textsubscript{2} and O3-CoO\textsubscript{2} are shown in figure \ref{fig4}d-i. 

\subsection{Cluster model calculations}

The single cluster configuration interaction calculations are carried out with the \textsc{Quanty} script language on cobalt-centered CoO\textsubscript{6} octahedral clusters \cite{haverkort_quanty_2016}. For the ground state calculations, only the low-energy Co 3d and O 2p states were considered, with 46 fermionic states. The Co 3d and O 2p onsite energies and the hopping parameters, including the d-p and p-p inter-atomic interactions within the cluster, were extracted from the tight-binding Hamiltonian obtained with \textsc{Wannier90} using a homemade python script. Due to the octahedral approximation, small but non-zero off-diagonal d-d and p-p inter-orbital interactions were also included (order of 10 meV). In second quantization formalism, the single electron Hamiltonian $H^{(1)}$ of the cluster can be written as: 
\begin{align}
H^{(1)} = &\sum_{i}{\epsilon_{i}d^{\dagger}_i d_i}
+ \sum_{i \neq j}{t_{i,j}d^{\dagger}_i d_j} + \nonumber\\
&\sum_{i}{\epsilon_{i}p^{\dagger}_i p_i} +
\sum_{i \neq j}{t_{i,j}p^{\dagger}_i p_j} +
\sum_{i,j}{t_{i,j}(d^{\dagger}_i p_j + h.c.)}
\end{align}
with $\epsilon$ and $t$ being the on-site and hopping energies read from \textsc{Wannier90}, respectively, $d^{\dagger}$ and $d$ ($p^{\dagger}$ and $p$) the creation and annihilation operators for the Co 3d shell (O 2p shell), and $i$ and $j$ spin-orbital indices. The full cluster model Hamiltonian is:
\begin{equation}
H^{cluster} = H^{(1)} + H_{dd}^U + H_d^{SO} - H_{dc}^{AMF}
\end{equation}
where $H_{dd}^U$ the Coulomb repulsion between two Co 3d electrons, $H_d^{SO}$ the spin-orbit coupling for the Co 3d shell, $H_{dc}^{AMF}$ is the double counting correction in the around mean field approximation. The $H_{dd}^U$ term is expanded in spherical harmonics leading to the following expression for 3d shell:
\begin{equation}
H_{dd}^U = \sum_{k=0,2,4} {F_{dd}^k H^{F^k}} 
\end{equation}
with $F^k$ the radial parts, taken as Slater integrals, and $H^{F^k}$ the spherical counterpart of the k-th order pole. The spin-orbit coupling is defined as:
\begin{equation}
H_d^{SO} = \xi_{d} \sum_i {(l_i \cdot s_i)}
\end{equation}

The Slater parameters (with a 80 \% scaling) and spin-orbit coupling for Co$^{3+}$ and Co$^{4+}$ configurations were taken from the \textsc{Crispy} library of Hartree-Fock values and are shown in table T1 in the Supplementary Information \cite{retegan_crispy, Suppl_Inf}. The monopole part (k=0) of the Coulomb interaction is related to $U_{dd}$ by \cite{haverkort_multiplet_2012}:
\begin{equation}
F_{dd}^0 = U_{dd} + 2/63 (F_{dd}^2 + F_{dd}^4 )
\end{equation}

To remove the correlation interaction accounted by the PBE potential at the DFT level, a double counting correction term  was subtracted from the total Hamiltonian. It consists of a mean field version of $ H_{dd}^U$ where all two-particles parts are replaced with the expectation values of the Co 3d density matrix obtained by the exact diagonalization of the DFT Hamiltonian $H^{(1)}$.

On top of the as-defined Hamiltonian $H^{cluster}$, a configuration-interaction model is taken into account to introduce the ligand-to-metal charge transfer $\Delta$, defined as $\Delta = E (\ket{{Co_{3d}}^n {O_{2p}}^{36}}) - E (\ket{{Co_{3d}}^{n+1} {O_{2p}}^{35}})$,
with n = 6 and 5 for LiCoO\textsubscript{2} and CoO\textsubscript{2}, respectively. From this definition and including the Coulomb interactions, the trace average $\epsilon_d$ and $\epsilon_L$ of the Co 3d and O 2p on-site energies are then set to the following definitions:
\begin{equation}
\epsilon_d = \frac{36\Delta - n(n+71)U_{dd}/2}{n+36}
\end{equation}
\begin{equation}
\epsilon_L = n \frac{n(n+1)U_{dd}/2 - \Delta}{n+36}
\end{equation}
The ground state of $H^{cluster}$ is found by exact diagonalization in the restricted active space of n between 6 (5) and 8 for LiCoO\textsubscript{2 }(CoO\textsubscript{2}) and with a convergence limit of 10\textsuperscript{-10} eV. The three lowest energy eigenstates were always computed, however we found the energy difference between the lowest and the second-lowest to be in the order of 0.1 - 0.5 eV. We therefore used only the lowest eigenstate for expectation values and spectroscopy calculations.

The photoemission spectral functions are computed from the negative imaginary part of the Green’s function: 
\begin{equation}
    G(\omega) = \bra{\Psi_{G.S.}} T^\dagger \frac{1}{\omega - H + i \Gamma/2} T \ket{\Psi_{G.S.}}
\end{equation}
where $\Psi_{G.S.}$ is the lowest-energy many-body eigenfunction of $H^{cluster}$, $T$ is the transition operator, $H$ the Hamiltonian for the final state, and $\Gamma=0.1$ eV the core-hole lifetime. In case of Co 3d and O 2p direct (inverse)  photoemission spectra, $T$ is the annihilation (creation) operator on the respective shell. For the Co 2p spectra, isotropic incident X-rays with a 45° angle to the surface were included, corresponding to our instrumental configuration. The O 2p spectra were averaged and weighted to give the 2:1 multiplicity of the formula unit; a scaling factor of 1/10 was also taken into account for the Co 3d and O 2p photoelectron cross sections at 1.5 keV \cite{leckey_subshell_1976}. The calculated spectra were broadened by Gaussian and Lorentzian functions with full width at half maximum of 0.7 and 0.3 eV, consistently with our experimental resolution. The restrictions for the Co 3d spectra calculations were obtained by those of the ground state calculation by removing or adding one electron for the direct or inverse photoemission, respectively (e.g. the Co 3d spectra for LiCoO\textsubscript{2} were simulated in the restricted active space between 3d\textsuperscript{5} and 3d\textsuperscript{7} configurations). The restrictions for the O 2p spectra calculation were the same as for the ground state calculation.

\begin{table}[hbt!]
\caption{\label{tab:table2} Slater and spin-orbit parameters used for the cluster model calculations. All values are expressed in eV.}
\begin{ruledtabular}
\begin{tabular}{c c c}
                &       LiCoO$_2$   &   CoO$_2$ \\
\colrule
$F_{dd}^2$      &       10.130      &   10.910  \\
$F_{dd}^4$      &       6.333       &   6.858   \\
$\xi_{Co_{3d}}$ &       0.074       &   0.082    \\
$F_{pd}^2$      &       10.130      &   10.910  \\
$F_{pd}^4$      &       6.319       &   6.835   \\
$G_{pd}^1$      &       4.758       &   5.220  \\
$G_{pd}^3$      &       2.707       &   2.973   \\
$\xi_{Co_{2p}}$ &       9.746       &   9.746    \\
\end{tabular}
\end{ruledtabular}
\end{table}

While for the Co 3d and O 2p spectra $H = H^{cluster}$, the model Hamiltonian for Co 2p spectra calculation $H^{XPS}_{Co_{2p}}$ was built upon $H^{cluster}$ to include the Co 2p core states interactions, namely the core spin-orbit coupling and the Coulomb interaction with the Co 3d shell, expressed by direct and exchange terms of the multipole expansion:
\begin{align}
    H^{XPS}_{Co_{2p}} = &H^{cluster} +
    \xi_{Co_{2p}} \sum_i {(l_i \cdot s_i)} + \nonumber\\
    &F_{pd}^0 H^{F_{pd}^0} +
    F_{pd}^2 H^{F_{pd}^2} +
    G_{pd}^1 H^{G_{pd}^1} +
    G_{pd}^3 H^{G_{pd}^3} 
\end{align}
where for $\xi_{Co_{2p}}$, $F_{pd}^2$, $G_{pd}^1$, and $G_{pd}^3$ we used the Hartree-Fock values of the \textsc{Crispy} library (table \ref{tab:table2}) and $F_{pd}^0$ was related to $U_{pd}$, for which we assumed the common approximation of $U_{pd} = 1.2 U_{dd}$ \cite{bocquet_electronic_1996}. Again, the trace average of the onsite energies was set to the configuration interaction model, this time including the core-valence Coulomb interaction, giving:
\begin{equation}
    \epsilon_p = -n U_{pd}
\end{equation}
\begin{equation}
    \epsilon_d = \frac{36\Delta - n(n+71)U_{dd}/2 - 216 U_{pd}}{n+36}
\end{equation}
\begin{equation}
    \epsilon_L = \epsilon_d + n U_{dd} + 6 U_{pd} - \Delta
\end{equation}
The same angle, polarization, and broadening used for the valence band calculations were used for the core level spectra. The occupation of the 3d shell was restricted as for the ground state calculation. 

\section{Acknowledgments}
We thank J. Ast and S. Motellier for their support on XRD and ICP-Ms investigations. We acknowledge C. Secouard for providing the LiCoO$_2$ thin films used in this study. This work was supported by the “Recherches Technologiques de Base” program of the French National Research Agency (ANR) and by CEA FOCUS-Battery Program. The work was carried out at the platform of nano-characterization (PFNC).

\nocite{apsrev41Control}

\bibliographystyle{apsrev4-1}
\bibliography{PRL_bibliography}

\end{document}